# Induction of Forces Performed by Piezoelectric Materials

Elio B. Porcelli[1]* and Victo S. Filho[1]

[1]H4D Scientific Research Laboratory

São Paulo, São Paulo State, 04674-225, Brazil

*Corresponding Author: elioporcelli@h4dscientific.com

**Abstract**

We describe the phenomenon of generation of an external field of forces from piezoelectric materials subjected to the application of electric fields or mechanical stress. The piezoelectric materials are shown as being capable of producing induction forces in external objects and we conclude that the nature of the forces generated are not originated from the traditional interactions. Further we specifically assert that the generation of forces by the piezoelectric materials is ruled by the hypothesis of preexisting condition of generalized quantum entanglement between the molecular structure of the material bulk and the surrounding environment. In addition the widely spread coupling of the molecules with the environment can be manifested from the so-called direct effect or the converse effect in piezoelectric materials and this coupling is not intermediated by acoustic waves or electromagnetic fields. We show that the novel effect has a theoretical explanation consistent with the generalized quantum entanglements and the direction of the induced forces depends on either the direction of the mechanical force or the electric field applied in these materials.

**Keywords:** piezoelectric materials, generalized quantum entanglement, induction of forces

## 1. Introduction

As known, the piezoelectric effect (Cady, 1964) is the capacity of some materials like crystals and ceramics to generate and garner electric charges when is applied in them a mechanical stress (Jaffe, 1971). The phenomenon was discovered by brothers Curie in 1880 and it was named as piezoelectric effect due to the greek words piezo and piezein that mean push and to squeeze (or press), respectively. The piezoelectric materials can suffer two effects, that is, they can be subjected to the direct effect (the generation of electric charges when stress is applied) and also exhibit the reverse piezoelectric effect, that is, the generation of stress when it is applied an external electric field. Under mechanical stress, the materials suffer unbalance in the positive and negative charges in their bulk, which then results in an external electric field. If the process is reversed, an outer electrical field can both as stretch as compress the piezoelectric material.

In present days, such properties of piezoelectric materials are used in many applications involving the production and detection of sound, fabrication of sensors (Gautschi, 2002), production of filters and other high-frequency applications (Newnham, 1980), generation of high voltages and electronic frequency, production of ultrasonic, actuators (Uchino, 1986; Karpelson, 2012), motors (Uchino, 1996) and many other applications. We here propose that relevant new applications can be performed considering some emergent properties of such piezoelectric materials.

A special use of these materials refers to their polarized molecular structure when subjected to voltage or to (mechanical) contact forces, respectively applied when either the direct effect or the reverse effect takes place. We here report that the piezoelectric materials present an unexplored property that refers to the generation of an external field of forces in the environment that is not due to the electric field generated in the direct process. We also assert that the main physical agent for that new application is the quantum entanglement between the polarized molecules and the external particles in the environment (some of which are part of macroscopic objects). This weak coupling of the huge set of molecules and the environment can macroscopically be uncovered when the voltage is applied to the piezoelectric material. The coupling between the polarized molecules and all other external particles is performed according to the concept of Generalized Quantum Entanglement (GQE), as early described in similar effects already analyzed in the literature (Porceli, 2015a; Porcelli, 2016a; Porcelli, 2016b; Porcelli, 2016c). When a voltage or a mechanical force is applied to piezoelectric materials such as either a quartz crystal or a PZT ceramic, an amount of momentum is transferred from the polarized molecules to the external particles in the case. The momentum direction depends on the direction of the electric field or the mechanical force applied. This particular attribute of a piezoelectric material allows a considerable induction of force in other external objects, although the effect is not visually detectable due to the weakness of the forces generated. External objects can suffer the induction of forces by the piezoelectric materials independent of their constitution. The induction can affect all kind of objects or particles and this is not performed for example by acoustic waves or electromagnetic fields (electromagnetic interactions can only affect electrical charged particles). Such an induction is related to the wide quantum coupling among the particles predicted by the Generalized Quantum Entanglement (GQE) concept. Such an anomalous effect has a similar counterpart in the case of superconductor materials operating under high external fields and or rotation that produces motion in external objects as for instance a simple pendulum (Podkletnov, 2001; Podkletnov, 1997; Podkletnov, 2003). This latter anomaly also has no traditional explanation and the Generalized Quantum Entanglement can also explain it, as reported in the literature (Porcelli, 2016d). Equivalent effects explainable by this theory are Biefeld-Brown effect (Porcelli, 2016a) and anomalous forces produced by magnetic cores (Porcelli, 2016c).

As here reported, the intensity of the induction depends directly on the intensity of the homogeneous electric field (voltage) or the mechanical force applied. Other dependence is related to the piezoelectric parameters of the materials microscopically defined by the internal quantity of polarized molecules and their collective geometry.

The divergence of the induction of forces is determined by parameters such as the shape of the bulk of piezoelectric material, the homogeneity of the internal polarized molecular density and the homogeneity of the electric field or mechanical force applied. Considering this points, the induction may affect external objects placed in high distances from the position of the piezoelectric material bulk. In other words, the space geometry of the induction can be focused or not, but this can be adjusted accordingly.

In the next section, we describe the experimental procedure and the device set in order to detect such anomalous forces generated by the piezoelectric materials under operation.

## 2. Description of the Experimental Procedure

As commented in the last section, it is well known that piezoelectric materials find wide use through their main property named "converse effect" to convert electrical energy to mechanical energy where the application of an electrical field creates deformation in the crystal and mechanical force. The other main property named "direct effect" is related to the conversion from mechanical energy to electrical energy where the application of a mechanical force produces voltage. Surprisingly, these materials can be used as inductors of external force and such an induction is not intermediated by acoustic waves or electromagnetic fields, but is caused by collective displacement of the internal polarized molecules when either direct or converse effects takes place and their mutual coupling with the external environment induces the generation of anomalous forces.

Here the innovative feature is that there is a coupling between these polarized molecules and the external environment via widely existing quantum entanglements. Properly adjusting some parameters such as the intensity of the electric field or mechanical force applied in the material and also the properties of the material, it is possible to control the force inducted in the external targets placed in any medium. With basis on such a principle we constructed a device or invention that will now be explained in more detail in the figures described next in this work. The experimental procedure to detect the anomalous forces induced by the piezoelectric body and characterize the device is reported from now on. It is relevant to say that the figures next are purely diagrammatic and not drawn to scale.

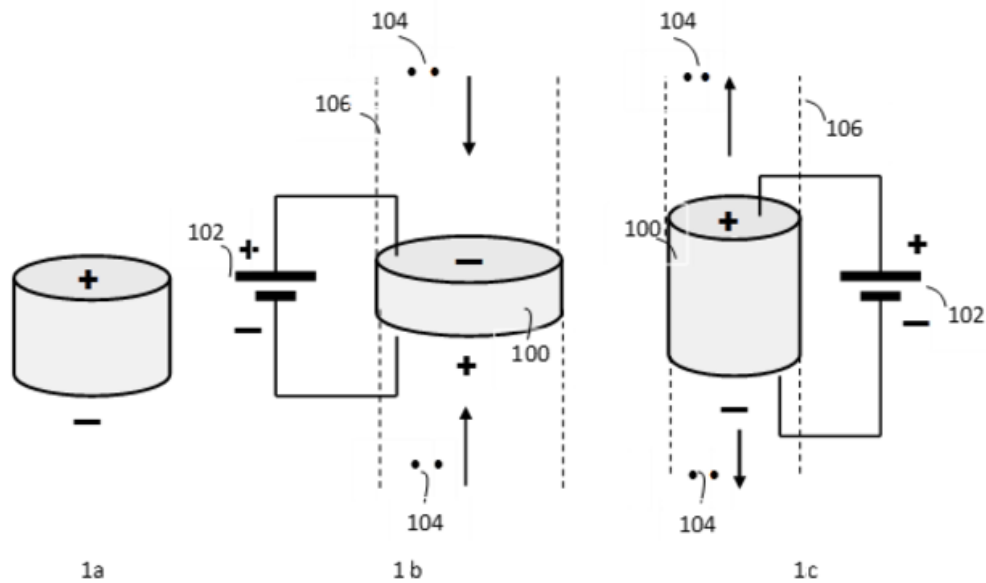

Figure 1. Diagrammatic scheme of a cylindrical piezoelectric material mounted in order to study the anomalous effect in the reverse case. In the left, in Fig. 1a we have the cylinder without the application of voltage, in which no significant forces appear in the medium. In the middle (Fig. 1b), a vertical attraction appears in the external targets when it is reversely polarized by the battery. In the last one, at the right we see in Fig. 1c that a vertical repulsion in the external targets appears when this same cylinder is directly polarized by the battery. In the scheme, the meaning of the numbers are: 100 – piezoelectric body; 102 – battery; 104 – outer particles; 106 – projected area.

In Figure 1, it is diagrammatically drawn a geometrical projection of the force induction that emerges from the piezoelectric cylinder making a vertical attraction in the external targets when it is reversely polarized by the battery such as shown in Figure 1b and making a vertical repulsion in the external targets when this same cylinder is directly polarized by the battery such as shown in the figure 1c. Negligible induction is observed when it is not applied a voltage in the piezoelectric cylinder, such as shown in the figure 1a. As seen in the scheme drawn in Figure 1b the geometrical projection of the force induction emerging from the cylinder vertically attracts external targets. In this condition, the polarized molecules of the cylinder behave such as electric dipoles and are mutually re-oriented into the cylinder structure by the vertical electric field. The shape of the piezoelectric cylinder is compressed in the vertical direction. The molecules are coupled with other external particles in the environment via generalized quantum entanglements. Considering this attribute, the external particles placed through the projection of the circular area of the cylinder undergo an attraction diagrammatically shown by arrows following the re-orientation of the internal polarized molecules. When the piezoelectric cylinder is directly polarized by the battery (Figure 1c), a vertical repulsion diagrammatically shown by arrows in the external targets takes place. The shape of the piezoelectric cylinder is extended in the vertical direction. In this condition, the external particles placed through the projection of the force induction of the cylinder follows the re-orientation of the internal polarized molecules in the outer direction from the circular faces considering the coupling via generalized quantum entanglements. The projection of the force induction when the piezoelectric cylinder is polarized (reversely or directly) can be collimated without attenuation with the distance apart of the cylinder area depending on the geometry of the cylinder, the uniformity of the density of internal molecules and the parallelism of the electric field lines applied.

In Figure 2, we see a diagram to represent the direct effect in a piezoelectric body. As before, a very weak and negligible induction (which can be associated to noises or interferences on the measurement devices) is observed if it is not applied a mechanical force in the piezoelectric cylinder, such as shown in the figure 2a. In figure 2b, it is shown a geometrical projection of the force induction that emerges from the piezoelectric cylinder making a vertical attraction in the external targets when it is mechanically compressed and at right making a vertical repulsion in the external targets when this same cylinder is extended mechanically, such as shown in the figure 2c. In that figure, the scheme of the geometrical projection of the force induction that emerges from the piezoelectric cylinder is drawn with the indication of a vertical attraction in the external targets when it is mechanically compressed (figure 2b). In this condition, the piezoelectric cylinder is compressed by a contact force that acts at the point of contact with other object. In fact, objects do not actually touch each other; rather contact forces are the result of the electrical interactions of the electrons at or near the surfaces of the objects. These interactions propagate molecule by molecule from both surfaces to the core of the cylinder. In this way, the internal molecules of the cylinder are reoriented into its structure following the direction of the contact force. These molecules are coupled with other external particles in the environment via generalized quantum entanglements. Considering this attribute, the external particles placed through the projection of the circular area of the cylinder undergo an attraction diagrammatically shown by arrows following the reorientation of the internal polarized molecules. When the piezoelectric cylinder is mechanically extended, such as shown in the Figure 2c, a vertical repulsion diagrammatically shown by black arrows in the external targets takes place. In this condition, the piezoelectric cylinder is extended by a contact force that acts at the point of contact with other object. As mentioned before, objects do not actually touch each other; rather contact forces are the result of the electrical interactions of the electrons at or near the surfaces of the objects. These interactions propagate molecule by molecule from both surfaces to the core of the cylinder. The internal molecules of the cylinder are reoriented in the outer direction from the circular faces of the piezoelectric cylinder. The external particles placed through the projection of the circular area of the cylinder are repulsed considering its coupling with the cylinder molecules via generalized quantum mechanics according to the direction of the contact force. The projection of the force induction when the piezoelectric cylinder is mechanically compressed (or extended) can be collimated without attenuation with the distance apart of the cylinder area depending on the geometry of the cylinder, the uniformity of the density of internal molecules and the alignment of the acoustic shock waves propagating into the piezoelectric material.

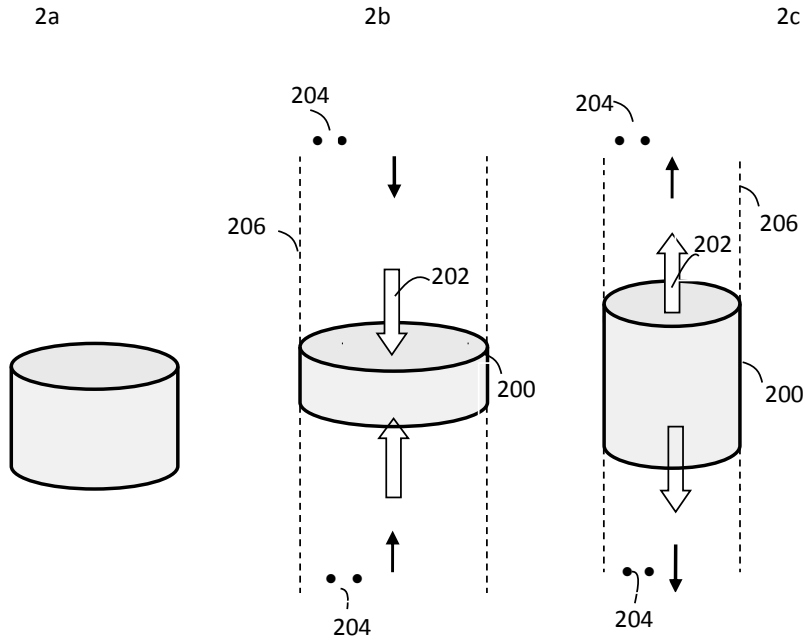

Figure 2. Schematic diagram of the experimental procedure for the direct case by considering a cylindrical piezoelectric body. In Fig. 2a (to the left), we see that there is no effect without the application of any mechanical force. In the Fig. 2b, we have a compression tension (white arrows), which induces the external forces (black arrows) in direction to the cylinder. In the last case, in Fig. 2c (to the right), we see the opposite action, that is, a traction force is performed on the cylinder (white arrows) generating repulsive forces (black arrows). In the scheme, the numbers indicated mean: 200 – piezoelectric body; 202 – compression or traction forces; 204 – external particles; 206 – projected area.

In Figure 3, we show the curve of the mechanical admittance Y or mobility M according to the frequency F in the vibration mode of the piezoelectric material. The intensity of the force induced by the piezoelectric body follows directly this curve of mechanical admittance and the point of higher intensity is placed in the resonance frequency $F_r$ of the piezoelectric material. In that figure, it was considered a continuous contact force applied in case of the direct effect such as shown in the Figures 2b and 2c or a force generated by a continuous electric field in case of the converse effect such as shown in the Figures 1b and 1c so far. The "continuous" word means that the action is invariable in the time. The piezoelectric materials have a characteristic behavior when subject to variable forces in the time. In case of converse effect where the application of a variable electrical field in the time creates mechanical deformation also variable in the piezoelectric material, the conversion rate from the electrical to the mechanical energy follows graphically a curve of mechanical admittance or mobility M (Figure 3). The analogue behavior can be found in case of direct effect where the conversion rate from the mechanical energy to the electrical energy when a mechanically variable contact force in time is applied. The peak of the curve of mechanical admittance (or mobility M) can be graphically found for some frequency named $F_r$ (resonance frequency). For this value of frequency, the maximum energy conversion rate can be found accordingly.

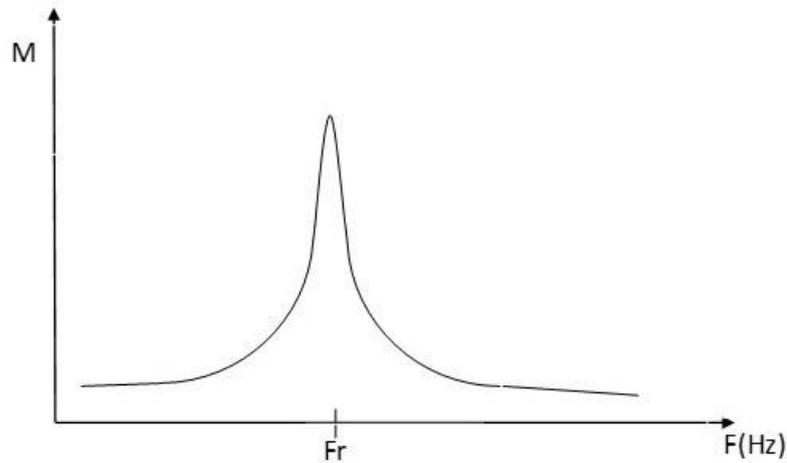

Figure 3. Plot of the admittance Y or mobility M as function of the frequency F in the vibration mode of the piezoelectric material.

In Figure 4, it is shown the diagram in which a specific use of our device shows the direct effect for a piezoelectric ceramic disc PZT4. The application of the mechanical contact force variable in time perpendicular to the flat surfaces creates an induction force at the same direction. The direction of the induction force is shown in the area A emerging from the piezoelectric disc. The projection of the forces is crossing an acoustic and electromagnetic barrier (shown as the wall in the Fig. 4). The barrier can block any acoustic or electromagnetic signal or at least reduce its intensity in order to be undetectable. The physical property of the projection of induction of force allows it can cross any barrier and, considering this, the accelerometer on a side of the barrier in the figure opposite to the piezoelectric disc is crossed by the induction force can detect its intensity accordingly. The induction force detected is variable in the time. The loudspeaker indicated in figure 4, with 4 Ohm of electrical impedance, makes a necessary mechanical vibration of the piezoelectric disc where it is coupled. The loudspeaker is electrically connected via two wires in an audio amplifier. The audio signal generator of sinusoidal wave produces the electric signal to be amplified by the power amplifier.

The setup shown in the Figure 4 can be used for metrology considering the actual difficulties to mark two or more points that need to be geometrically linked by a straight line in the huge and massive structures where it can not be crossed easily by light laser beam, other electromagnetic signals or acoustic waves. Currently the methodology for this procedure is expensive, inaccurate and time consuming, considering that many external sensors are used around the structure where the measurements are made indirectly.

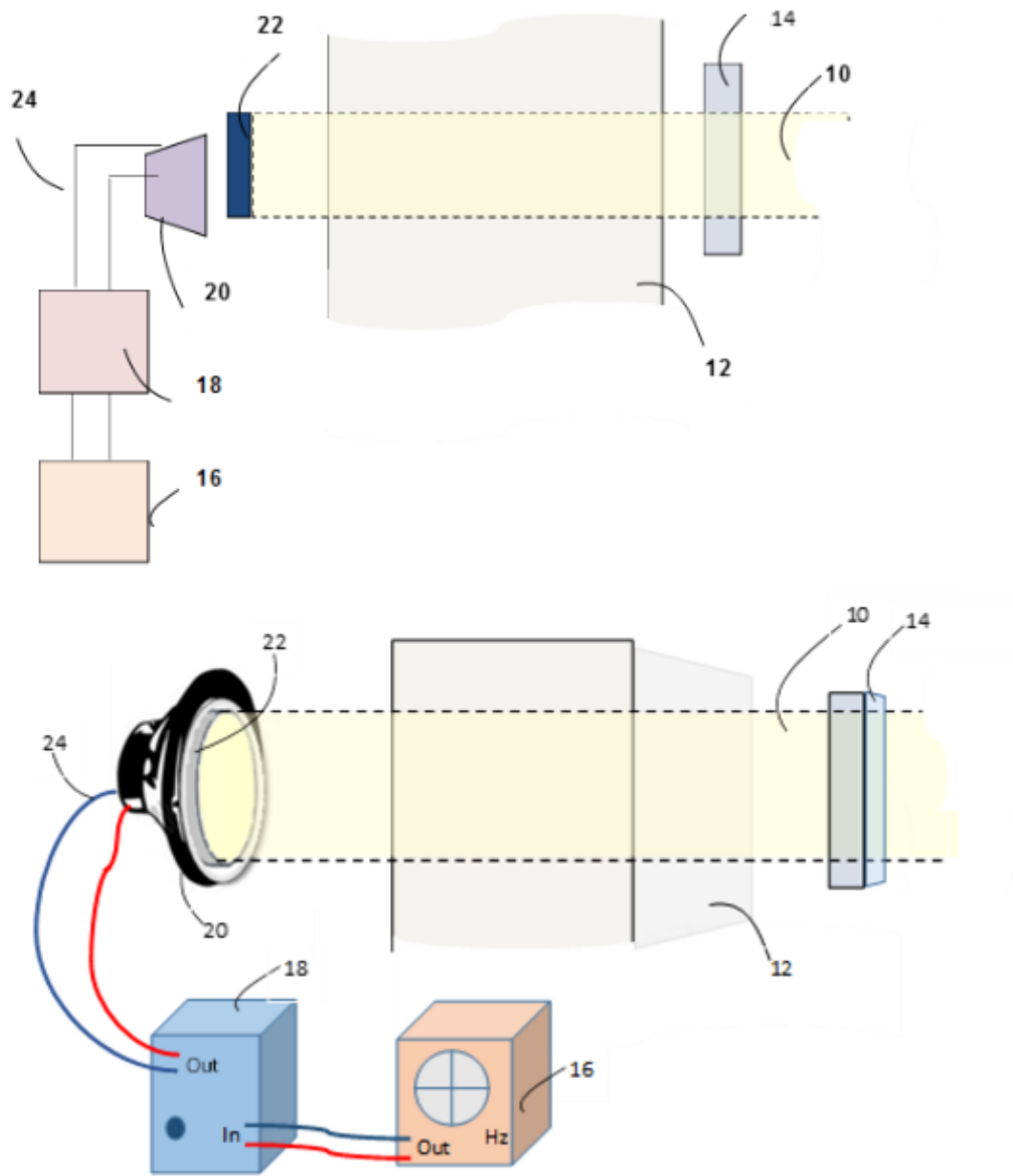

Figure 4. A scheme of the device that produces induced forces in the close environment from the operation of the piezoelectric body. In the diagram shown, we can identify the following components: 10 – induction force; 22 – piezoelectric disc; 12 – barrier wall ; 14 - accelerometer; 24 – two wires of connection; 18 – audio amplifier; 16 - audio signal generator, 20 - loudspeaker.

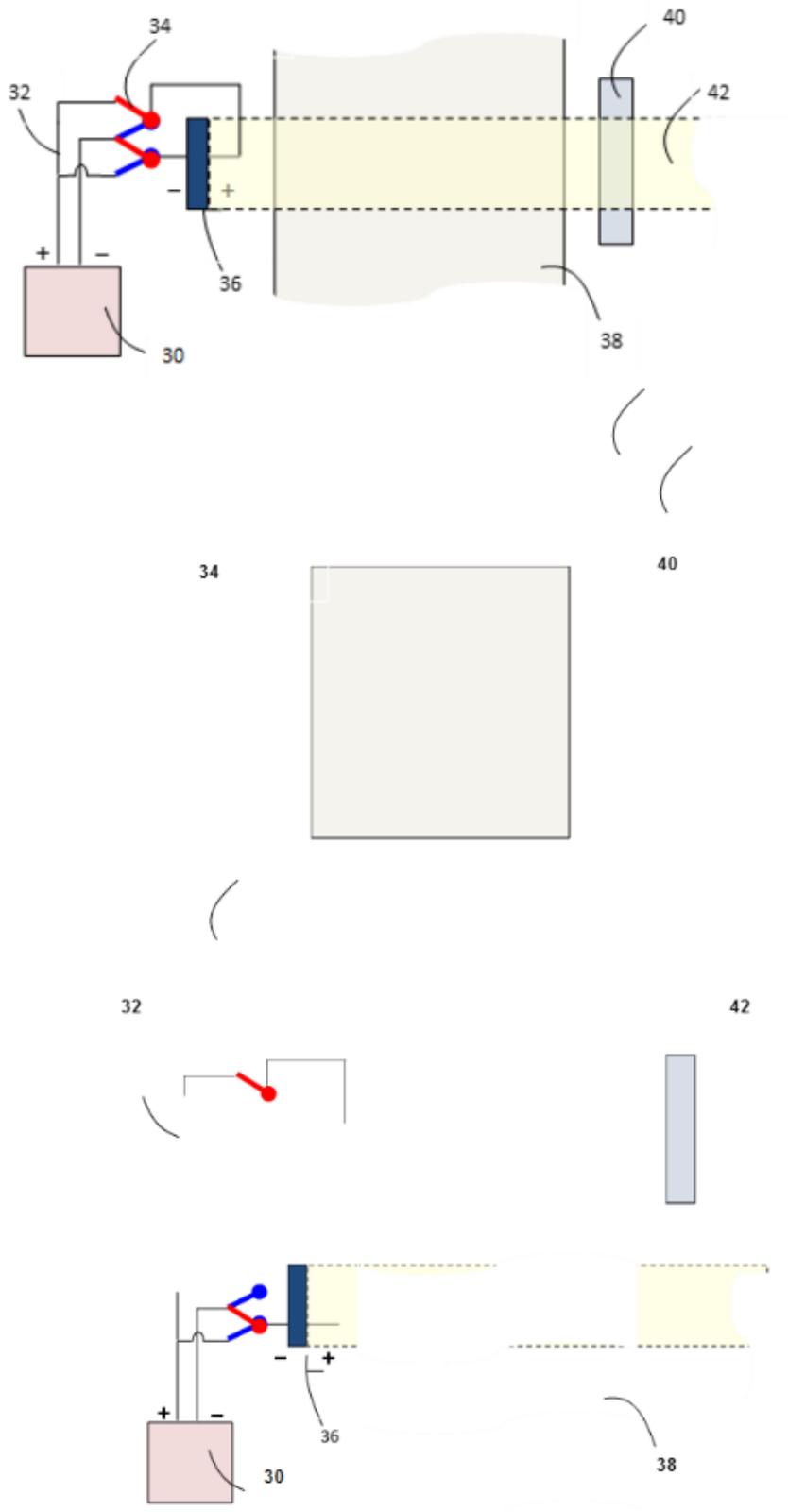

Figure 5. A scheme of the device invented in order to induce forces in the close environment by using the direct property of piezoelectric materials. The numbers indicated in the scheme represent: 36 – piezoelectric disc; 42 - induction force; 38 – barrier wall; 40 – accelerometer; 30 - power supply; 32 - connection wires; 34 – twin switches.

In Figure 5, we also show a diagram in which a specific use of our invention is related to the direct effect for a piezoelectric ceramic disc PZT4. The projection of the induction forces showed in Figure 5 is emerging from a piezoelectric disc and crossing an acoustic and electromagnetic barrier, that is, the wall indicated in Figure 5. The physical property of the projection of induction forces allows it can cross any barrier and, considering this, the accelerometer indicated in figure that is crossed by the induction can detect its intensity accordingly. The high voltage power supply provides a DC voltage via the wires drawn in Figure 5 for both circular faces of the piezoelectric ceramic disc PZT4 with 5cm diameter and 2.5mm thickness. The two twin switches seen in Figure 5 can be adjusted accordingly to polarize the piezoelectric ceramic disc. In the red position, the switch connects the positive pole of the power supply in the positive face of the piezoelectric ceramic disc and it connects at the same time the negative pole of the power supply in the negative face of the piezoelectric ceramic disc. The shape of the disc is expanded and it projects a repulsive force induction. In the blue position, the switch connects the positive pole of the power supply in the negative face of the ceramic disc and it connects at the same time the negative pole of the power supply in the positive face of the piezoelectric ceramic disc. In this condition, the shape of the disc is compressed and it projects an impulsive force induction.

As early reported from Figure 4, the setup shown in the Figure 5 can be also used for metrology considering the actual difficulties to mark two or more points that need to be geometrically linked by a straight line in the huge and massive structures where it cannot be crossed easily by acoustic or electromagnetic waves (so called local interactions). Currently the methodology for this procedure is made considering that many external sensors are used around the structure where the measurements are indirectly made.

**2.1 Description of the Converse Mode Experiment**

In the following, we describe the instruments used in the experimental work, the real experimental setup conceived in order to measure the effect and experimental procedure adopted to avoid possible interferences and noises and implement an effective technique of measurement of the anomalous forces early described.

The list of the main devices used in both experiments (PZT4 disc converse and direct mode) comprises:

1. Non Brand PC computer running a tone generator software (96 KHz maximum) to generate an audio signal to its RCA output;
2. Active Subwoofer system manufactured by Clone model 11137, frequency range from 150Hz to 18 KHz, Power 8W RMS, input AV voltage 127 or 220 VAC (device used like audio amplifier to excite the 4 Ohms loudspeaker according to the signal received from the PC computer via RCA input);
3. 4 Ohm Loudspeaker – frequency range from 150Hz to 18 KHz;
4. PZT4 disc provided by ATCP do Brasil, 5cm diameter, 2.5 mm thickness, 913.2 KHz resonance frequency;
5. Non brand High-voltage power supply model AT 30 KV, output range 0~27 kV DC, voltage resolution 100 V Ripple 150V, Standard output current 0.3 mA @ 15kV, Power 20W, Input AC voltage 127 or 220 VAC;
6. USB accelerometer Gulf Coast Data Concepts Model X6-2, educational proposes, stand-alone operations, 3 axes accelerometer, 2g or 6g acceleration range in each axis.

In Figure 6, it is shown a diagram of the experimental setup where the measurements of the converse mode for the PZT4 disc takes place. The base of the cone (diaphragm) of the loudspeaker supports the PZT4 disc where they are both mechanically coupled with their symmetry axes aligned each other according to the vertical direction. The acoustic and electrostatic insulation box made by Styrofoam covered by a thin aluminum foil encloses all this mentioned setup. An acoustic vibration of the PZT4 on vertical direction is taken in place when an amplified sinusoidal signal is supplied to the loudspeaker via audio amplifier. The audio amplifier receives the signal from the computer running a special software where the frequency, wave shape and amplitude can be adjusted accordingly. It was generate an audio signal with three frequencies: 500, 5000 and 10000 Hz. The accelerometer is placed apart in other different support (shelf made by glass) than PZT4 disc (both supports are not coupled mechanically in order to avoid vibration and noise exchanging) and enclosed in the acoustic and electrostatic insulation box also made by Styrofoam covered by a thin aluminum foil. All electrostatic shields (aluminum foils) are connected to the ground of the laboratory. The "z" axis sensor of the accelerometer is aligned according to the symmetry axis of PZT4 disc in the vertical direction. The projection area of the PZT4 disc transgress (illuminates) the accelerometer; this last one is 0.62 m above (and away) position than the first one. When the audio signal is turned on, the accelerometer reads a variation of the gravity acceleration induced by a force originated by the PZT4 disc in reverse mode.

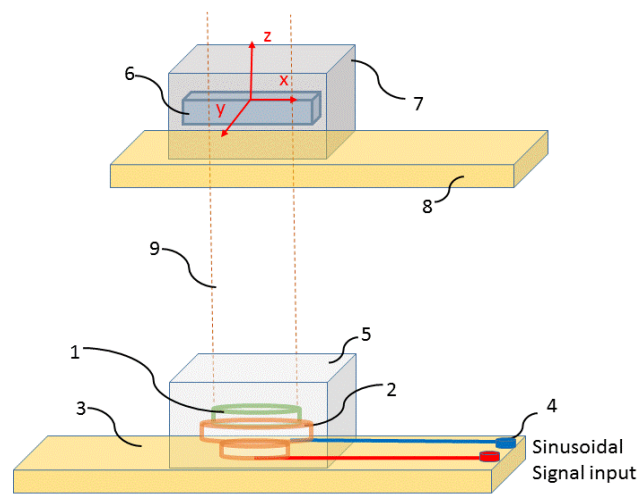

Figure 6. A scheme of the setup of converse mode experiment. In the diagram shown, we have the following components: 1 - PZT4 disc; 2 - Loudspeaker; 3 - Table shelf; 4 - Electrical contacts of the loudspeaker (sinusoidal signal input); 5 - Acoustic and electrostatic insulation box (Styrofoam covered by thin aluminum foil); 8 - Accelerometer support shelf; 9 - Projection of the PZT4 disc area.

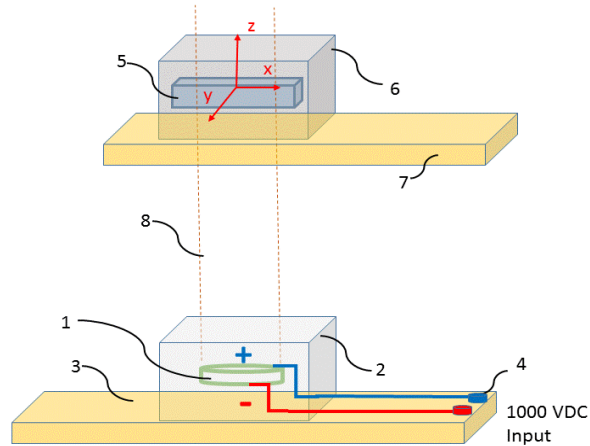

Figure 7. A scheme of the setup of the direct mode experiment. In the diagram shown, we have the following components: 1 - PZT4 disc; 2 - Acoustic insulation made by box of Styrofoam covered by thin aluminum foil; 3 - Accelerometer support shelf of glass; 4 - Electrical contacts of the PZT4 disc (1000 VDC input); 5 - Accelerometer; 6 – Acoustic and electrostatic insulation box (Styrofoam covered by aluminum foil); 7 - Accelerometer support shelf; 8 - Projection of the PZT4 disc area.

The analysis of the accelerometer measurements were made in two periods of time after it is switched on; the first one was from 1th to 30th second and the following one was from 31th to 60th second. The audio signal was turned on between 30 and 60 seconds after switched on the accelerometer. Because of such a procedure, the first step of the analysis of accelerometer measurements was made when the audio signal was switched off and the second step was made when the audio signal was switched on.

The difference between two periods ("on" and "off") of average measurements indicated an gravity acceleration variation. The accelerometer was configured to provide high resolution sampling, measuring 320 samples per second.

**2.2 Description of the Direct Mode Experiment**

In this case, the procedure is very similar to the latter described in the last section. In figure 7, it is shown a diagram of the experiment where the measurements of the direct mode for the PZT4 disc take place. The acoustic and electrostatic insulation box made by glass covered by a thin aluminum foil encloses the PZT4 disc with its symmetry axis positioned in the vertical direction. A horizontal shelf made by wood supports the whole mentioned setup. Two aluminum insulated stripes connect electrically each circular side of the PZT4 disc in order to the high voltage (1000 VDC) can be applied on it. The PZT4 disc is electrically polarized (one circular side is positive and other side is negative) and its shape dimension can variate in the vertical direction (stretches or contract) depending on the polarization of the voltage applied. The accelerometer is placed apart in other different support (iron shelf) than PZT4 disc (both supports are not coupled mechanically in order to avoid vibration and noise exchanging) and enclosed in the acoustic and electrostatic insulation box also made by Styrofoam covered by a thin aluminum foil. All electrostatic shields (aluminum foils) are connected to the ground of the laboratory. The "z" axis sensor of the accelerometer is aligned according to the symmetry axis of PZT4 disc in the vertical direction. The projection area of the PZT4 disc transgress (illuminates) the accelerometer, this last one is 0.6 m above (and away) position than the first one. When the 1000 VDC high voltage is turned on, the accelerometer read a variation of the gravity acceleration induced by a force originated by the PZT4 disc in direct mode.

The analysis of the accelerometer measurements was made in two periods of time after it is switched on; the first one was from 1th to 30th second and the following one was from 31th to 60th second. The 1000 VDC high voltage was turned on between 30 and 60 seconds after switched on the accelerometer. Because of such a procedure, the first step of the analysis of accelerometer measurements was made when the 1000 VDC high voltage was switched off and the second step was made when the voltage was switched on.
As in the converse case, the difference between two periods ("on" and "off") of average measurements indicated an gravity acceleration variation. The accelerometer was configured to provide high resolution sampling, measuring 320 samples per second.

In case of experiments with PZT4 disc in converse and direct mode, all shields (boxes and shelves) were implemented in order to reduce or even eliminate the possibility to the induced force be related to any acoustic or electrostatic (electromagnetic) cause, that is, so called local interactions. It were added more flat barriers like wood and glass plates between the emitter (PZT4 disc) and receptor (accelerometer) without apparent attenuation of the induced force magnitude reinforcing the argument for nonlocal interactions as the cause of the induced force. The induced force became undetectable when the accelerometer suffered a displacement of a few centimeters in the "x" or "y" sensor directions to outside position of project area of the PZT4 disc. It was not well detectable the attenuation of the induced force according to the "z" axis direction length considering the limited space available for the setup in the laboratory. The accelerometer measurements of the magnitude of the induced force via gravity acceleration variations are in accordance with the theoretical forecasts showed in this work. These results encourage the patent application (Porcelli, 2015b).

### 3. Theoretical Explanation and Calculation of the magnitude of nonlocal force induction
The magnitude of the nonlocal force induction in external objects or particles generated by the piezoelectric material in converse or direct mode depends on the intensity of the electric field applied for its inner collective molecular reorientation but a binding electrical intermolecular (and interatomic) interaction can act against this reorientation. In this way the magnitude of the nonlocal force induction depends directly on the compression or distention (external mechanical force applied) of the piezoelectric material such as a cylinder or disc in its axis direction in case of converse mode. The dependence relates to the external voltage applied on the piezoelectric material in case of direct mode. The magnitude of the nonlocal force induction depends inversely on a macroscopic parameter of elasticity of the material named "Young Modulus" for converse and direct mode.

A piezoelectric parameter (Pereira, 2010) $g_{33}$ for PZT ceramics (the most usual piezoelectric material in the market - Lead Zirconate Titanate) indicates the relationship between three quantities (V, F and T). V is the voltage externally applied for direct mode or internally generated in case of converse mode. F is the mechanical force externally applied for converse mode or internally generated in case of direct mode. T is the thickness of axis length of the cylinder or disc. The following formula represents this mentioned relationship:

$g_{33}$ = V . T / F.

Considering the previous knowledge about the voltage V applied or generated and about some parameters of PZT ceramic such as $g_{33}$ and T, it is possible to calculate the magnitude of mechanical force applied or generated using the relationship mentioned before:

F = V . T / $g_{33}$.

The strain S can be calculated considering the formula

S = F / A,

where S is the strain in the piezoelectric cylinder; F is the mechanical force as mentioned before and A is the circular area of the cylinder.

A parameter named "deformation" has not dimensionality in terms of physical quantity and it measures the rate of physical deformation of the piezoelectric material when a voltage (direct mode) or mechanical force (converse mode) is applied on it. This parameter D can be calculated by the following formula:

D = S / Y,

where S is the strain in the piezoelectric material (cylinder or disc), Y is the Young modulus and D is the deformation.

The parameter D can be multiplied to the value of the mechanical force in order to calculate the magnitude of the nonlocal force induction as the formula shown:

f = F . D,

where f is the magnitude of the nonlocal force induction, F is the mechanical force and D is the deformation parameter. In case of null voltage applied in the piezoelectric cylinder, there is a negligible nonlocal force induction generated considering the null deformation in this condition, such as shown in the Figure 1a for direct mode and Figure 2a for converse mode.

The nonlocal force induction can be attractive or repulsive depending respectively on compression (or contraction) or distention of the piezoelectric material such as shown in figures 1b and 1c for converse mode and figures 2b and 2c for direct mode. The signal of the values of the parameters F and D determines if the nonlocal force is attractive or repulsive according to the inertial reference of the piezoelectric material. In this case, the attraction or repulsion of the external particles has no dependence with their attributes such as electrical charge, for example.

In case of time-varying regime, the mechanical admittance (or mobility M) for some frequency of mechanical oscillation (or electric field oscillation) is deeply linked with the natural oscillation frequency of the polarized molecules of the piezoelectric material coupled mutually via intermolecular and interatomic electric interactions. The polarized molecules are also coupled with external particles via generalized quantum entanglements. Considering this, the intensity of the force induction can be calculated by the formula

$f = F_a \cdot F / F_r$,

where f is the intensity of nonlocal force induction, $F_a$ is the variable mechanical force in time existing in the piezoelectric material for direct or converse mode, F is the frequency oscillation of the force applied and $F_r$ is the resonance frequency of the piezoelectric material. This formula is valid for a value of the frequency oscillation of the force applied F lower than the resonance frequency $F_r$ ($F \leq F_r$). The intensity of the force induction calculation for some specific sample of piezoelectric material must be made by using the values (F and $F_r$) from the curve of the mechanical admittance or mobility M such as shown in the graphic of the Figure 3. This curve can be obtained by measurements of the sample applying different values of oscillation frequencies.

In case of converse mode experiment and considering the maximum power provided by the audio amplifier, the total amplitude of the sinusoidal (mechanical) contact force applied in the symmetry axis direction of the PZT4 disc by the loudspeaker is $7.33 \times 10^{-3}$ N for a frequency as 500 Hz according to the average of measurements made directly by the accelerometer coupled to both PZT4 disc and loudspeaker for this purpose.

The contact force $F_a$ is applied to the circular surface of the piezoelectric ceramic disc. The intensity of the induction force can be calculated according to the formula $f = F_a \cdot M / M_r$ as shown in the graphical representation of the Figure 3. Regarding that point, M is the admittance (or mobility) for the (mechanical) contact force $F_a$ applied with a frequency F and $M_r$ is the admittance (or mobility) for the (mechanical) contact force when the resonant frequency $F_r$ takes place.

The admittances M and $M_r$ respectively related to the frequencies F and $F_r$ can be obtained from the specific curve of admittance (or mobility) for the piezoelectric ceramic disc PZT4, as shown in the figure 3.

In the case where the frequency F is much lower than the frequency $F_r$ ($F << F_r$), the rate $F / F_r$ can be used instead of $M / M_r$ for a good approach.

Considering this information, we can calculate the magnitude of the nonlocal force induction by the formula

$f = F_a \cdot (F / F_r) = 7.33 \times 10^{-3} \cdot [500 / (913.4 \times 10^3)] = 4 \times 10^{-6}$ N .

Considering the intensity of the induction force f and considering an accelerometer mass "$M_a$" equal to $39.3 \times 10^{-3}$ Kg, the acceleration "a" measured by the accelerometer can be calculated by Newton formula, as follows:
$a = f / M_a = 4 \times 10^{-6} / (39.3 \times 10^{-3}) = 1.02 \times 10^{-4}$ m / s$^2$

This value related to the acceleration is in accordance with the average for the values measured by the accelerometer, considering also the standard deviation.

The intensity of the induction force is weak but it is enough to be detected by the accelerometer. This mentioned setup is preliminary but the parameters, features and new materials can be improved in order to generate a strong induction force in the external objects for general purposes.

The best performance (maximum intensity of induction force) can be achieved when it is generated a high power oscillation with a frequency with the same value than the resonant frequency (F=$F_r$).

The piezoelectric ceramic disc considered in this setup has a 5 cm diameter, 2.5 mm thickness, $g_{33}$ parameter equal to 0.02292 V m / N and $Y_{33}$ parameter equal to $6.2 \times 10^{10}$ N / m$^2$. It is considered a maximum DC voltage applied between its circular faces equal to 1000 V.

In case of direct mode, this information allows us also to calculate the intensity of the attractive or repulsive nonlocal force induction measured by the accelerometer.

The modulus of the force for the 1000 V voltage applied in the piezoelectric ceramic disc can be calculated using the formula $F = V \cdot T / g_{33}$ as showed before as follows:

$F = 1000 \times 2.5 \times 10^{-3} / 0.02292 = 109.1$ N.

The second step is to calculate the strain S using the formula $S = F / A$:

$S = F / A = 109.1 / (1.964 \times 10^{-3}) = 55564.17$ N/m$^2$,

where A is the circular area of the disc.

The next step is to calculate the physical deformation D using the formula $D = S / Y_{33}$:

$D = S / Y_{33} = 55564.17 / (6.2 \times 10^{10}) = 8.967 \times 10^{-7}$.

Finally, the nonlocal force induction can be calculated according to the calculation as shown:

$f = F \cdot D = 109.1 \times 8.972 \times 10^{-7} = 9.78 \cdot 10^{-5}$ N

For this magnitude of the nonlocal force induction and considering the accelerometer mass "$M_a$" equal to $39.3 \times 10^{-3}$ Kg, the acceleration "a" measured by the accelerometer can be calculated by Newton formula, as follow:

$a = f / M_a = 9.78 \times 10^{-5} / (39.3 \times 10^{-3}) = 2.49 \times 10^{-3}$ m/s$^2$.

As before, this value of acceleration "a" is also according to the average value of the measurements made by the accelerometer. The magnitude of the nonlocal force induction is also weak but enough to be detected by an accelerometer. As already mentioned, such a setup is preliminary but the parameters, features and new materials can be improved in order to generate a strong induction force in the external objects for general purposes.

The best materials for direct mode need to have a small value for $g_{33}$ parameter and a high value in terms of electrical insulation in order to support a high voltage application.

From such an invention (Porcelli, 2015b), we got a method for using piezoelectric materials or devices as inductors of force in other external objects comprising generalized quantum coupling between its internal polarized molecules and external particles or objects placed in the beam environment.

**4 - Final Considerations**

It is well known that the piezoelectric materials in converse mode can emit acoustic waves or electromagnetic radiation in case of direct mode so that a force can be transferred at the distance to outer bodies. Up to now all present devices or procedures usually use these mentioned properties.

This work shows that a new application of piezoelectric materials in both converse and direct mode occurs: the induction of force at the distance is also transferred to outer bodies considering a physical property such as the preexisting condition of generalized quantum entanglement between all involved parts of the system, that is, emitters (piezoelectric materials) and sensor receivers.

This new application can be used on remote sensing, that is, acquisition of information about an object or phenomenon without making physical contact with the object and thus in contrast to on-site observation or metrology (science of measurement) using a piezoelectric material (e.g. PZT4 ceramic) as signal emitter and the sensor (e.g. accelerometer) as receiver.

A trivial application is to mark two points into a solid structure (huge or small) in order to build a tunnel or hole. Remote sensing or metrology can be also performed via acoustic waves (e.g. ultrasonic microwaves) or electromagnetic radiation (e.g. radio signal) generated by piezoelectric materials but there is some disadvantage situation in comparison than using the induction of force at the distance via quantum entanglements. The main disadvantage is related to the attenuation or even insulation of acoustic or electromagnetic signals depending on the propagation medium considering that this effect is absent for quantum entanglements.

It is remarkable that the effect of the preexisting condition of generalized quantum entanglements is extremely weak, but during the period of converse mode or the direct mode, a myriad of internal molecular electric dipoles transfer their collective momentum variation (still weak but at least better measurable during this transitory state) to outer bodies like sensors. The characterization of this mentioned property was performed in laboratory accordingly and the magnitude of the induction of nonlocal force induction can be calculated only using the known parameters of the piezoelectric materials.

It is also relevant to consider that the effect here described can be enhanced by different types of piezoelectric materials (Newnham, 1980). For instance, techniques in the fabrication have been described as successful in the increasing of the magnitude of the piezoelectric coefficients. The influence of crystal symmetry, macro-symmetry, and interphase connectivity have been used to explore piezoelectric composites that have been prepared with higher piezoelectric coefficients. Further some materials have been developed by mixing volatilizable plastic spheres and PZT powder that give excellent piezoelectric voltage coefficients. Large voltage coefficients were also obtained from piezoelectric composites. So, we can conclude that there is a wide margin for research in the field so that we can produce devices with more significant and intense forces generated.